\documentstyle[11pt,epsf]{article}

\input psfig
\def\beq{\begin{eqnarray}}
\def\eeq{\end{eqnarray}}
\def\bea{\begin{eqnarray*}}
\def\eea{\end{eqnarray*}}

\def\centeron#1#2{{\setbox0=\hbox{#1}\setbox1=\hbox{#2}\ifdim
\wd1>\wd0\kern.5\wd1\kern-.5\wd0\fi
\copy0\kern-.5\wd0\kern-.5\wd1\copy1\ifdim\wd0>\wd1
\kern.5\wd0\kern-.5\wd1\fi}}
\def\ltap{\;\centeron{\raise.35ex\hbox{$<$}}{\lower.65ex\hbox{$\sim$}}\;}
\def\gtap{\;\centeron{\raise.35ex\hbox{$>$}}{\lower.65ex\hbox{$\sim$}}\;}

\def\singleandthirdspaced{\baselineskip=\normalbaselineskip\multiply
    \baselineskip by 130\divide\baselineskip by 100}
\def\singlespaced{\baselineskip=\normalbaselineskip}

\newcommand{\newc}{\newcommand}
\newc{\qbar}{{\overline q}}
\newc{\Kahler}{K\"ahler }
\newc{\deltaGS}{\delta_{\rm GS}}
\begin{document}
\begin{titlepage}
\begin{flushright}
{\large hep-th/0002047 \\ SCIPP-2000/04\\
}
\end{flushright}

\vskip 1.2cm

\begin{center}

{\LARGE\bf Towards a Solution of the Moduli Problems of String Cosmology}

\vskip 1.4cm

{\large  Michael Dine}
\\
\vskip 0.4cm
{\it Santa Cruz Institute for Particle Physics,
     Santa Cruz CA 95064  } \\

\vskip 4pt

\vskip 1.5cm

\begin{abstract}
There are at least two serious moduli problems in string
cosmology.  The first is the possibility that moduli dominate the
energy density at the time of nucleosynthesis.  The second
is that they may not find their minima all together.  After
reviewing some previously proposed
solutions to these problems, we propose another:
all of the moduli but the dilaton sit at points of enhanced
symmetry.  The dilaton has a potential similar to those
of racetrack models; it is very massive and its dynamics do not
break supersymmetry.   The dilaton is able to find the minimum
of its potential because the energy is dominated by non-zero
momentum modes.  This energy need not be thermal.  The effective
potential for the dilaton is quite different from its flat space
form.  If certain conditions are satisfied, the dilaton
settles into the desired minimum; if not, it is forced
to weak coupling.

\end{abstract}

\end{center}

\vskip 1.0 cm

\end{titlepage}
\setcounter{footnote}{0} \setcounter{page}{2}
\setcounter{section}{0} \setcounter{subsection}{0}
\setcounter{subsubsection}{0}

\singleandthirdspaced


\section{Introduction:  Moduli in Cosmology}

Moduli are ubiquitous in string models, but they are almost
certainly not present in nature.  Most ideas for fixing the moduli
still leave over approximate moduli.
In pictures where
supersymmetry is relevant to understanding the hierarchy, there
are some approximate moduli which are associated with supersymmetry breaking
(e.g. in models of gluino condensation\cite{gauginocondensation}, Kahler
stabilization\cite{kahlerstabilization},
or the racetrack picture\cite{racetrack,iy,dsracetrack}).
In scenarios with large
or warped dimensions, there are approximate moduli associated with
the size of the extra dimensions.  These approximate moduli are simultaneously
interesting and dangerous for cosmology.   They are interesting
because they are candidate inflatons\cite{gaillard,banksinflation}, and could play a
role in generating the baryon asymmetry\cite{ad}.  They are problematic
because they can easily carry two much energy\cite{bkn}, or simply
fail to find their minima all together\cite{bs}.  The first of
these difficulties is usually called the ``cosmological moduli
problem," the second the ``Brustein-Steinhardt problem."

Several solutions to the cosmological moduli problem have been
proposed.  The moduli might simply be much more massive than one
might have expected from considerations of hierarchy; if their
masses are of order $10$'s of TeV, or their interactions
somewhat stronger than expected from naive considerations, their
decays
can restart nucleosynthesis.  The baryon number can be produced
either directly in their decays, or through the Affleck-Dine
mechanism\cite{yanagidaetal}.  Late inflation\cite{lateinflation} (and thermal inflation
\cite{thermalinflation}) have been
discussed as solutions.

One particularly simple possibility is that
of ``maximally enhanced symmetry"\cite{drt,
dns,yukawa}.  This is the idea that all of
the moduli transform non-trivially under unbroken (or slightly
broken) symmetries.  The main difficulty with this idea is
connected with
what we will refer to as the dilaton, $S$, that modulus (or more
generally moduli) which controls the values of the standard model
gauge couplings.  It seems unlikely that the dilaton would have an
enhanced symmetry at a point where the gauge couplings are small
(some ideas for how a weak coupling might emerge were discussed in
\cite{dns}).  So one might consider, instead, the possibility that
the dilaton does not sit at an enhanced
symmetry point; this point is determined by
supersymmetry-preserving dynamics at some
high scale.  Any other moduli sit at enhanced symmetry
points.  Provided that the mass of the dilaton is greater than a
few $10$'s of TeV (in the mechanisms to be discussed below the
natural mass scale is much larger), this would solve the
cosmological moduli problem:  the dilaton would decay early; the
other moduli could naturally start out near their minima\cite{dsracetrack}.

This still leaves the Brustein-Steinhardt problem, and this will
be the focus of our attention in what follows.  The question is:
why should the dilaton end up anywhere near the correct vacuum.
In most pictures for the origin of the dilaton potential, the
potential is extremely steep.  This applies, in particular, to
both the racetrack models\cite{racetrack,dsracetrack} and to models of
Kahler stabilization\cite{kahlerstabilization}.  As a result,
if one assumes that the
zero mode of the
dilaton dominates the energy density, one
finds that the field inevitably overshoots the minimum, except
for very special initial conditions.

This problem is troubling, but a number of mechanisms have been
suggested which might mitigate it.  Horne and Moore\cite{hornemoore}
noted that the
dilaton moduli space has finite volume.  As a result, there is a
finite probability that the system will start out with a suitable
initial coordinate and velocity so as to end up in the correct
vacuum.  Banks et al\cite{banksetal} have
made several other points concerning the Brustein-Steinhardt
problem.  First, they note that there may be modifications to the
Kahler potential which affect the steepness of the potential.  As
we will see, this is relevant to both the Kahler stabilization and
the racetrack schemes.  On the other hand, we will
see that the required modifications of the Kahler potential
are not very plausible.  Essentially, it is necessary to fine
tune the Kahler potential over a significant volume of the field
space.

Second, and most importantly, Banks et al point out that it is not consistent
to focus simply on the zero modes.
We will review this
argument, and pursue its implications.
The energy stored in non-zero modes might be thermal or not.
In either case, there is an effective potential for the
dilaton, which one can think of as arising from the usual
coupling constant dependent corrections to the free energy
and/or the kinetic terms of the various fields.  If the system
is thermal, the thermodynamic free energy
is a function of the couplings, $\Omega(T,g^2)$.  If the coupling
is dynamical, and if the dilaton mass is small compared to
the temperature (which it is for temperatures below
$M_p$), this is just the dilaton potential.  Even if the
energy is non-thermal, and carried by non-zero (or
zero) modes of some field $\Phi$ (which
might be the dilaton itself), there will be corrections to
the energy of the system in powers of the coupling constant,
which again constitute a potential for the dilaton. In either
case, at weak
coupling, this potential is likely to be far larger than
the non-perturbative potential.  The minimum
of this potential will not coincide with $S_o$, that of the zero
temperature, flat space theory, so the dynamics is quite different
from that considered in \cite{bs}.  The zero coupling limit
can be a minimum or a maximum of the potential.  If it is a minimum,
the system may be driven to weak coupling.  If it is a maximum
(i.e. if the potential tends to zero from below) the system
may well be set gently into the ground state.  As we will
see, even if the weak coupling point is a minimum of the potential,
under plausible conditions, the system finds its way to the true
minimum.  If these conditions are not satisfied, the system
is driven to weak coupling.

If the system is truly in thermal equilibrium, one can be rather
definite, and we will consider this case as well.
At very weak coupling, one can compute the potential.  However,
one expects the true minimum to lie at a point where weak
coupling methods are not reliable.  For gauge
groups typical of the racetrack
scheme, we will see that perturbation theory is not reliable, and
we need to make some assumptions about the form of the potential.
With plausible assumptions, the system does indeed find its ground
state.  But it is also quite possible for the system to move
continually to ever weaker coupling.

In non-supersymmetric scenarios, such as the large radius and
warped compact dimension pictures which have been developed
recently\cite{largedimensions,precursors,rs}, it is harder to
make
definite statements.  One doesn't
have quite as detailed a picture of the underlying theory in this
case (e.g. one doesn't usually know much about the moduli
which determine the gauge couplings).
We will make some comments on these ideas in our
conclusions.

In sum, we conclude that the usual formulation of the
Brustein-Steinhardt problem is not appropriate.  The answer
to the question:  do
the moduli end up close to their true minima, depends
on the behavior of non-zero momentum modes.  With quite plausible
assumptions about the potentials and about the conditions
of the early universe, the system can find its stable ground
state.  Actually determining the course of events requires
far greater knowledge than we have at present about the
stabilization of the moduli and about the initial conditions
in the early universe.

\section{Why is the Dilaton Special?}

In the weakly coupled picture of the heterotic string theory, the
gauge couplings are
controlled by a field, $S$.  The universality of the $S$ couplings
can lead to coupling unification.  Given that we don't expect this
weak coupling picture to hold in any detailed way (if at all),
what field might be singled out?  Consider, for example, the
strong coupling limit of the heterotic string.  Here, the fields
$S$ and $T$ are both large, and $T$-dependent corrections to the
gauge couplings are important and can potentially spoil the
prediction of unification\cite{hv,wittency}.

In the enhanced symmetry picture which we are proposing,
it is natural to suppose that all but one linear combinations of moduli
are fixed and of order one in the appropriate fundamental units.
One linear combination is not.  Call this combination $S$.
Suppose it couples to both a ``hidden sector" gauge group, as well
as to the ordinary gauge groups.  Then, as we will discuss below,
one can consider various mechanisms which stabilize $S$ at a large
value, giving rise to weak couplings.  Whether couplings are
unified is a separate question.  This depends on whether it is
reasonable that the $S$ couplings are universal.  Similarly,
whether this picture is realized in any known approximate
moduli space is a question which requires investigation.
In the naive Horava-Witten picture, for example, both the
$S$ and $T$ moduli are large\cite{hv}, but for
special backgrounds there is unification.  These issues are under study
and will be discussed elsewhere.

For now, we will simply assume that it is sensible to focus on
one particular modulus, as suggested by the enhanced symmetry
picture we have proposed.  Our cosmological remarks below
are likely to be applicable to models in which several
moduli suffer from the Brustein-Steinhardt difficulty.

\section{Moduli Stabilization}

Two ideas for stabilizing the moduli have been seriously pursued.
Both focus on the puzzle:  how can a theory with no small
parameter generate a small gauge coupling.  Both assume that the
resolution to this question lies in holomorphy.  More precisely,
some dynamics generates a minimum for the dilaton (where we use
the term in the sense of the previous section) at large $S$.
Corrections to the superpotential are exponentially small in
$S$, but corrections to the Kahler potential may be large.

The first of these mechanism is known as Kahler stabilization.
Gaugino condensation or some similar non-perturbative effect is
supposed to generate a superpotential for the dilaton, which falls
to zero exponentially rapidly for weak coupling.  The
stabilization of the dilaton arises because of
properties of the Kahler potential.  It has been argued that this
is plausible, since one might expect large corrections to the
Kahler potential from its weak coupling value even when the
coupling is rather small.

The second of these mechanisms is known as the racetrack model.
Here, the idea is that the stabilization occurs through properties
of the superpotential.  For example, if one has two gaugino
condensates, one might expect the superpotential to look as:
\beq
W = A e^{-S/N_1} - B e^{-S/N_2}
.
\eeq
This superpotential has a stationary point at
\beq
S = {N_1 N_2 \over (N_1-N_2)} \ln(B/A).
\label{dilatonvalue}
\eeq
So if $N_1$ and $N_2$ are large, the coupling may be small.  In
order that any systematic analysis be possible, it is necessary
that $N_1 \approx N_2$\cite{dsracetrack}.  In addition, in
general, one does not expect to be able to compute the Kahler
potential in this scheme (though this may be possible under
certain circumstances).

An appealing version of the racetrack scenario has
been put forth in \cite{iy}.  These authors consider models with
$R$ symmetries, which can give rise to unbroken supersymmetry with
$W=0$ at the minimum.  From the perspective of the cosmological
issues which we have discussed in the introduction, this is a
particularly attractive possibility.  One might imagine that the
coupling is fixed at a small value, but in such a way that the
dilaton is quite massive, and rather harmless in cosmology.
In this model, one has a set of singlets, $X$, coupled to two
gauge groups with quantum modified moduli spaces, and nearly
identical beta functions.  The couplings include
\beq
{\cal L} = X (A~Q_1 Q_1 - B~Q_2 Q_2)
\eeq
where $X$ is one of the singlets
and $Q_1$ and $Q_2$ denoting matter in the two different gauge groups.
The potential, then, has the form
\beq
V = \vert A e^{-{S/N_1}} - B e^{-S/N_2} \vert^2.
\eeq
This has a supersymmetry-conserving minimum with $S$ given by
eqn. \ref{dilatonvalue} above.

\section{The Brustein-Steinhardt Problem}

Both the Kahler Stabilization and Racetrack scenarios are
characterized by potentials which fall exponentially with the
dilaton field.  At weak coupling, the Kahler potential for the
dilaton is
\beq
K(S,S^{\dagger})= -\ln(S+S^{\dagger})
\eeq
so the canonical field, $\phi$, is related to $S$ by
\beq
S = e^{\phi}.
\eeq
This means that the potential behaves
for very weak coupling as:
\beq
V(S) \sim {\rm exp}(-A e^{\phi})
\eeq
i.e. it is an extremely steep function of $\phi$ for large $\phi$.

\begin{figure}[htbp]
\centering
\centerline{\psfig{file=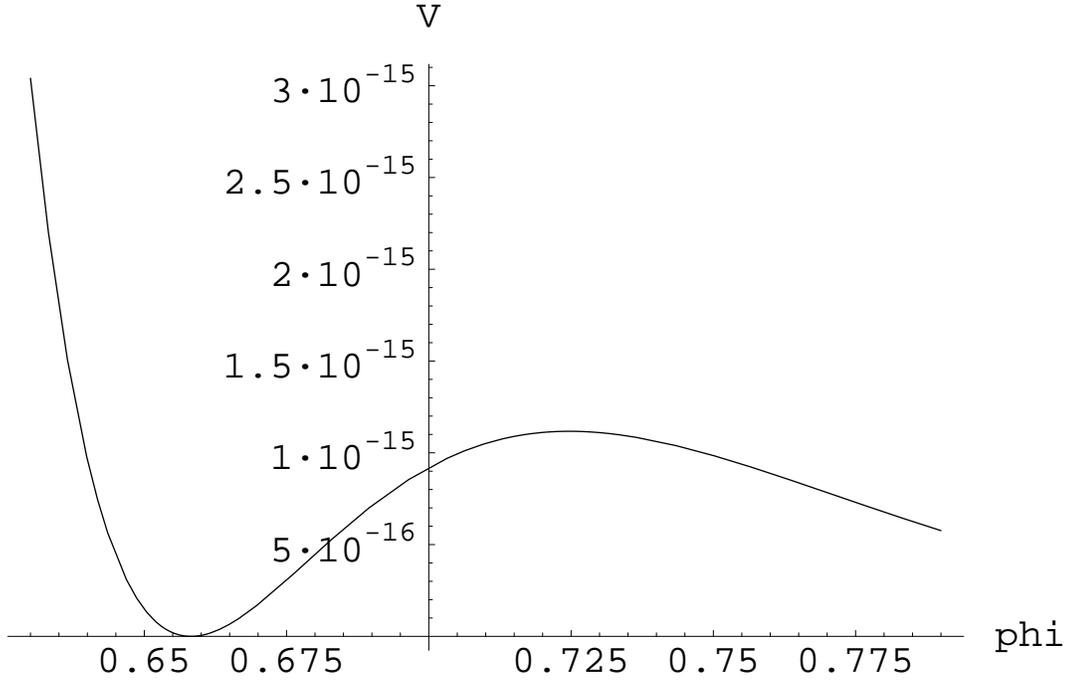,angle=-0,width=15cm}}
\caption{A typical racetrack potential (used in
calculations below).  Minimum is a small dimple.}
\label{dimple}
\end{figure}

In both of these proposals, the potential is already assumed to be
quite small at the true minimum.  So one has the situation
illustrated in fig. \ref{dimple}.  The potential is extremely steep
almost everywhere, with a tiny dimple at the minimum.  It is
natural to expect that the system overshoots.  To get some feeling
for the problem, first simply ignore the potential, and suppose
that $\phi$ dominates the energy density of the universe.  In this
case
$
p=\rho$
so the scale factor grows as
$R(t) \sim t^{1/3}.$
$\phi$
obeys the equation:
\beq
\ddot \phi + {1 \over t} \dot  \phi =0
,\eeq
with solution
\beq
\dot \phi = {c \over t}~~~~~~ \phi = \ln t + d.
\eeq
If one plugs this into the expression for $V$, one sees
that $V$ falls of exponentially fast with time, while the kinetic
energy falls off only as a power.   So the neglect
of the potential is consistent.  By the time the system
reaches the minimum, the potential energy is a tiny fraction
of the kinetic energy, and the system overshoots.  Numerical study
readily verifies that this is the case.

\section{Modification of the Kahler Potential}

The analysis above relied on the weak coupling form of the Kahler
potential.   In the case of Kahler stabilization, however, by
definition the Kahler potential is significantly different from
its weak coupling form near the minimum.  Similarly, in the case
of the racetrack scenario, one also expects significant
corrections to the Kahler potential\cite{dsracetrack}.  So it is
natural to ask\cite{banksetal} whether one could avoid the
Brustein-Steinhardt problem if the Kahler potential is
significantly different from its weak coupling form.  In this
section we explore this possibility.

One way to analyze this problem is to note that, if the
Brustein-Steinhardt problem is to be avoided, one should be in an
approximately slow-roll regime. Let us first formulate the conditions
for slow-roll in the context of a generic superpotential $W$ and
\Kahler\ potential $K$. For a Lagrangian of the form
\beq
{\cal L}=f(\phi,\phi^\dagger)\partial_\mu\phi^\dagger
\partial^\mu\phi-V(\phi,\phi^\dagger),
\eeq
the condition for slow-roll reads
\beq
{d\over d\phi}\left({1\over f}{dV\over d\phi}\right)\ll 3V.
\label{generalsr}
\eeq
In the supergravity framework, we have
\beq
V=\left|{\partial W\over\partial X}\right|^2 g^{-1}_{X\bar X}\ e^K,
\ \ \ f=K^{\prime\prime}.
\eeq
Let us denote $\bar V\equiv \left|{\partial W\over\partial X}\right|^2$.
We assume that it depends on the dilaton field $S$ and that
$g_{X\bar X}=1$. The condition (\ref{generalsr}) implies then
the following three conditions for $\bar V$ and $K$:
\beq
{\bar V^{\prime\prime}\over\bar V}\ll K^{\prime\prime},
\label{conda}
\eeq
\beq
{\bar V^{\prime}\over\bar V}\ll
\left({2K^\prime\over K^{\prime\prime}}-
{2K^{\prime\prime\prime}\over(K^{\prime\prime})^2}\right)^{-1},
\label{condb}
\eeq
\beq
1+{(K^\prime)^2\over K^{\prime\prime}}-
{K^\prime K^{\prime\prime\prime}\over(K^{\prime\prime})^2}\ll1.
\label{condc}
\eeq

Let us first investigate the case of the racetrack superpotential.
Equations (\ref{conda})-(\ref{condb}) can be rewritten then as
conditions on the \Kahler\ potential:
\beq
K^{\prime\prime}\gg1/N^2,
\label{corta}
\eeq
\beq
{(K^{\prime\prime})^2\over2K^{\prime}K^{\prime\prime}
-K^{\prime\prime\prime}}\gg1/N.
\label{cortb}
\eeq

At the minimum, $S={\cal O}(N)$. It is straightforward then to see
that with the weak coupling form, $K=-\ln(S+S^\dagger)$,
the condition (\ref{condc}) is satisfied but
the two conditions (\ref{corta}) and (\ref{cortb}) are not.
Allowing strong modifications of the \Kahler\ potential, such
as $K=S^n$ for any $n\geq2$ or $K=e^{aS}$ with $a$ positive or
negative, we still cannot satisfy the three conditions simultaneously.

One can understand that the problem of satisfying (\ref{condc}),
(\ref{corta}) and (\ref{cortb}) simultaneously is generic by
Taylor expanding around the minimum $S_0$. Take
\beq
K=a_0+a_1(S-S_0)+{1\over2}(S-S_0)^2+{1\over6}(S-S_0)^3.
\label{rttaylor}
\eeq
We are interested in finding a solution to the slow-roll condition
for $S$ that is within a few $N$ away from $S_0$, that is,
$\delta\equiv(S-S_0)/N={\cal O}(1)$. The $a_m$ coefficients
of eqn. (\ref{rttaylor}) should then have the following $N$ dependence:
\beq
a_m={\alpha_m}/N^m,
\eeq
where $\alpha_m$ are $N$-independent. One can write the three conditions
(\ref{condc}), (\ref{corta}) and (\ref{cortb}) in terms of the three
$\alpha_m$ and $\delta$. It becomes clear that fine tuning of the
$\alpha_m$ parameters is required to satisfy these conditions for
a given $\delta$. Even if we manage to find a solution for a given
$\delta={\cal O}(1)$, moving a distance $\Delta\delta={\cal O}(1)$
away from this point will reintroduce strong violations of the
slow-roll conditions.

We conclude that it is impossible to satisfy the slow-roll condition
for a racetrack superpotential and any non-singular \Kahler\ potential
for a finite range of $S$ that is a few $N$ away from the minimum.

Second, we would like to investigate the case of \Kahler\ stabilization.
The superpotential is of the form
\beq
W=e^{-aS}.
\eeq
Assume that $K$ can be expanded around the minimum $S_0$ of the scalar
potential $V$ as
\beq
\label{KKS}
K=K_0+K^\prime(S-S_0)+{1\over2}K^{\prime\prime}(S-S_0)^2
+{1\over6}K^{\prime\prime\prime}(S-S_0)^3.
\eeq
The scalar potential is then of the following form:
\beq
\label{VKS}
V=e^Ke^{-2aS}\left[\left|-a+K^\prime+K^{\prime\prime}(S-S_0)+{1\over2}
K^{\prime\prime\prime}(S-S_0)^2\right|^2{1\over K^{\prime\prime}
+K^{\prime\prime\prime}(S-S_0)}-3\right].
\eeq
Requiring that $V(S_0)=0$ and $V^\prime(S_0)=0$, we get
\beq
\label{zeroVZ}
K^{\prime\prime}={1\over3}(K^\prime-a)^2,
\eeq
\beq
\label{dVSZZ}
K^{\prime\prime\prime}={2\over9}(K^\prime-a)^3.
\eeq
Thus, setting $K^\prime-a$, we get $K^{\prime\prime}$ and
$K^{\prime\prime\prime}$. Combining analytical and numerical
searches, we were unable to find a solution of the slow-roll
equations for  $\delta\equiv(S-S_0)/N={\cal O}(1)$ that is not fine-tuned.

We conclude that it is highly unlikely that there exists a form
of the \Kahler\ potential that would both stabilize the dilaton
and satisfy the slow-roll conditions over a finite range of
$(S-S_0)={\cal O}(N)$.

\section{A Possible Solution:  The Role of Non-Zero Modes}

In the previous section, we concluded that modification of the
Kahler potential is unlikely to resolve the Brustein-Steinhardt
problem.  On the other hand,
the authors of \cite{banksetal} made an observation which both
sharpens the problem and is
likely crucial to its resolution.  They noted that the
focus on zero momentum modes alone is inconsistent.  Assuming the
behavior $R \propto t^{1/3}$ described above, the equation for the
non-zero momentum modes is:
\beq
\ddot \phi  + {1 \over t} \dot \phi + {k^2 \over R^2} \phi =0.
\eeq
This equation has the solution
\beq
\phi = {1 \over t^{1/3}} \cos{(3/2) t^{2/3}}.
\eeq
As a result, the energy density of the non-zero modes falls off as
$t^{4/3} \sim R^4$, i.e. more slowly than that of the zero modes
(which falls off as $1/t^2$), and just like that of radiation.

We have little insight as to what might be appropriate initial
conditions, but this result suggests some possibilities.  One is
to suppose that the field $\phi$ initially has a roughly thermal
distribution, and {\it dominates the energy
density}.  This is consistent with the $1/R^4$ falloff (of course,
the distribution does not have to be thermal to justify these
statements; it is only necessary that the energy density be
dominated by the non-zero modes of $\phi$).  This provides
additional damping, which, as we will see, can
appreciably slow the motion of  the field.  Just as important,
it also means that there is an additional
potential for the dilaton, a potential which can be much larger than
that due to the non-perturbative effects associated with gluino
condensation.  The point is that the action for the moduli
has coupling constant dependent corrections, i.e.
\beq
{\cal L}_{kin}= (\partial_{\mu}\Phi)^2(1+a g^2 + b g^4 + \dots).
\eeq
Averaging $(\partial_{\mu}\Phi)^2$ over the background,
generates a potential for the dilaton:
\beq
V(S) = \langle (\partial_{\mu}\Phi)^2 f(g^2) \rangle
\eeq
This potential
vanishes for large $S$ (small coupling).
Its behavior at large coupling, where we expect
$S_o$ lies, is unknown, and we can imagine many possibilities.

To see that the assumption of a non-zero momentum background significantly
alters the  picture, first ignore the potential all together,
as we did earlier.
The zero mode now
obeys the equation
\beq
\ddot \phi + {3 \over 2t} \dot \phi =0.
\eeq
This has solution
\beq
\phi = \alpha t^{-1/2} + \beta.
\eeq
In other words, the field creeps to some particular point.
Including the potential, it is reasonable to hope that the
system will track the potential, and eventually settle into the
correct minima.

\begin{figure}[htbp]
\centering
\centerline{\psfig{file=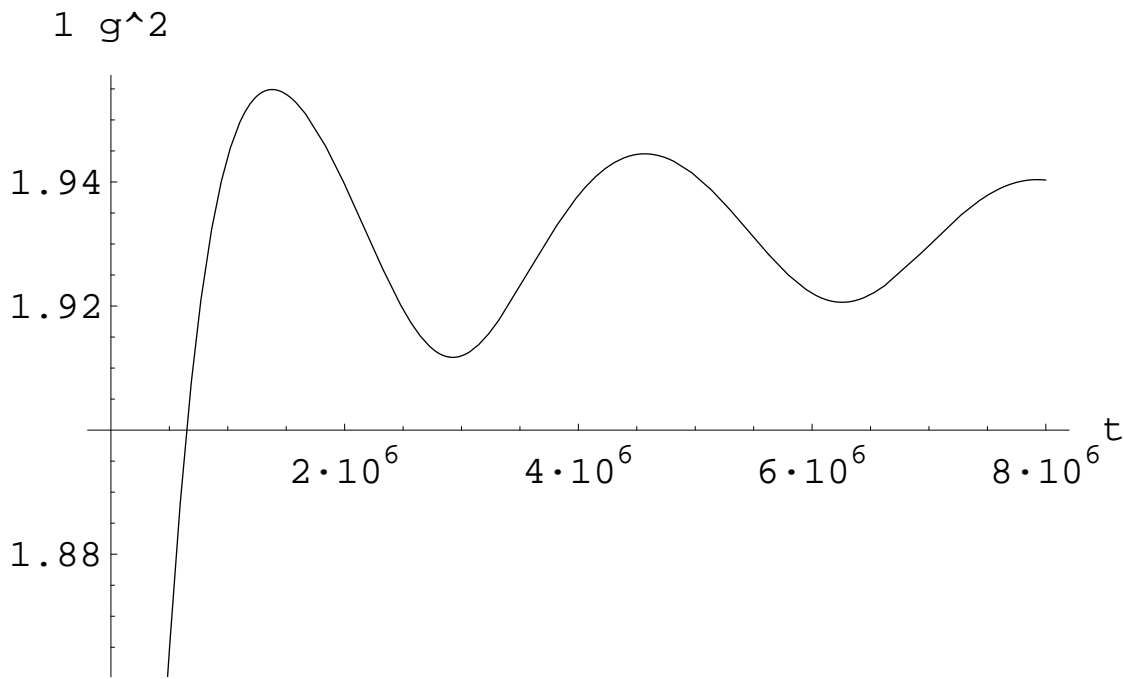,angle=-0,width=15cm}}
\caption{In the case that $b<0$, the system settles
nicely to its minimum.}
\label{negativecase}
\end{figure}

Now consider the problem with the potential,
including the large effects associated with the non-zero
energy.
Suppose that the potential tends to zero from below.
As a model, take the energy to have the form
\beq
V_{eff} =  b g^2+ V_{np}(g^2).
\eeq
Here $T$ is simply to be thought of as a parameter which
characterizes the energy of the system; it is not necessary that
the system be in thermal equilibrium, just that its energy
redshift like radiation.  If $b<0$,
the potential has a minimum for non-zero coupling,
resulting from the competition of the non-perturbative piece and
the additional, finite energy contribution.  If initially
the coupling is not too small, the finite energy contribution
dominates.  As this energy redshifts, the minimum moves to weaker
coupling; eventually, it will lie near $S_o$.  Provided $b$ is not too
small, the system tracks this minimum.  This can be seen in fig.
\ref{negativecase}, where we have taken $b=-.1$, and for $V_{np}$ we have taken
\beq
V_{np}= (e^{-{8 \pi^2\over 10 g^2}} -4  e^{-{8 \pi^2\over 11
g^2}})^2
\eeq
We have used the weak coupling form for the Kahler potential in
studying the motion of the field.
In the figure, one can see that the coupling evolves to its
minimum, and then oscillates about it, as expected.

\begin{figure}[htbp]
\centering
\centerline{\psfig{file=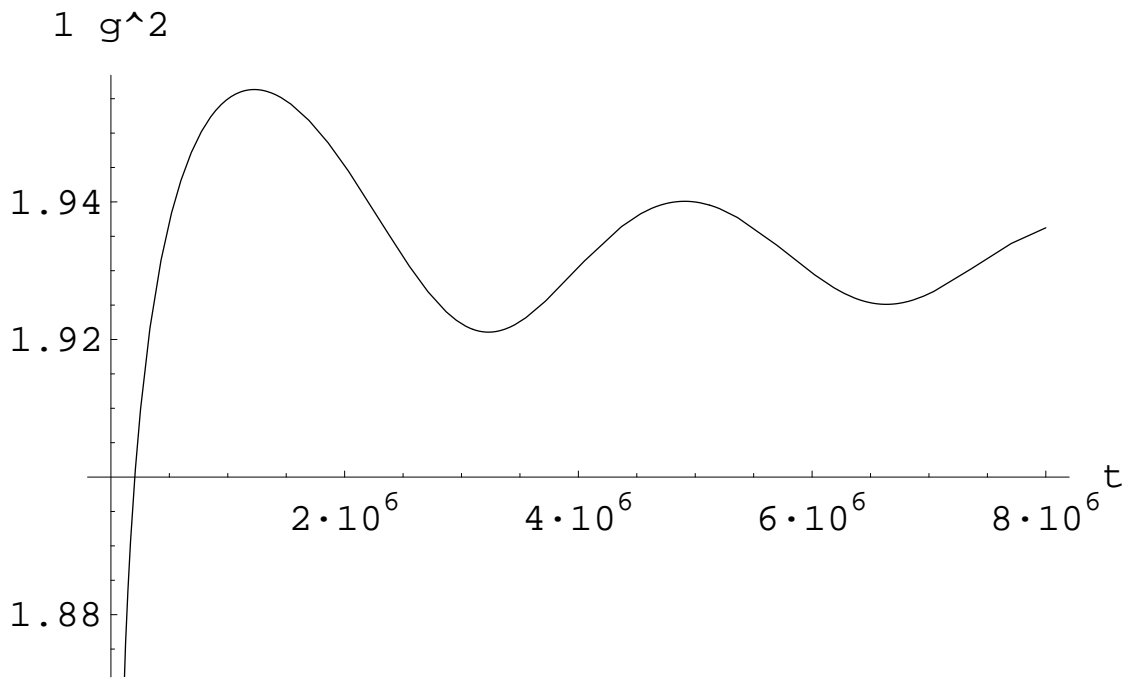,angle=-0,width=15cm}}
\caption{For small, positive $b$, the system also settles
to its minimum.  Here $b=0.02$.}
\label{positivecase}
\end{figure}

In the case that $b>0$, it is more difficult for the system
to find its minimum.  We find for this simple model that if
$b>0.02$, the system typically overshoots the minimum.  If
$b$ is smaller, however, the damping is adequate, and the system,
as in fig. \ref{positivecase}, settles into the minimum
near $S_o$ (in this case located at $S=1.91$).
This is compatible with crude estimates obtained by using the slow roll
approximation.

A second possibility is that the system really is in thermal
equilibrium.  In particular, the hidden sector gauge fields might
be in thermal equilibrium.  We need to ask what is the effective
action for $S$ in a thermal background.  But this is easy:  the
effective potential is precisely the free energy as a function
of coupling.
To understand this, note that it is appropriate to
think of the dilaton as changing adiabatically on the scale
$T^{-1}$.
As a result, we can integrate out the fast modes -- the
gauge bosons, gauginos, etc., to obtain an effective action for
$S$.  The potential just corresponds to setting $S = {1/g^2}$ to
a constant value, and computing the free energy
of the gauge system.  In other words, the potential for $S$ is just
the free energy of the gauge system.\footnote{This is different
than the results of \cite{steinhardt}, who argue that the
potential behaves as $1/g^2$, i.e. that it blows up at weak
coupling.  Their form is correct for coupling to
certain homogeneous scalar
figure configurations. Many of our other observations are similar
to theirs.}

The form of the free energy for such a
system is known.  For gauge group $SU(N)$ (without matter
fields, for simplicity) one has, for
example\cite{kapusta}
\beq
\Omega(g,T) = -{(N^2-1)\pi^2 \over 24} T^4 (1- {3 \over 8}{ g^2N
\over \pi^2} + { g^3 N^{3/2} \over \sqrt{2} \pi^3} + \dots).
\eeq
As expected, it tends to zero for small coupling.
This formula does not exhibit a minimum for larger coupling.
However, for $N=10$ or so, as expected in the racetrack picture,
the perturbation expansion has broken down at interesting
values of the coupling ($g \sim 1/2-1)$, and we might speculate
on possible behaviors.

The system might have no
minimum at all, except at zero coupling.
In this case, the problem is similar to that we considered
above, with a potential proportional to $g^2$
with a positive coefficient.  If one takes the perturbative
formula literally, this is roughly like $b \sim 1/3$,
for $N=10$, and so the system is likely to overshoot
the minimum.  On the other hand, given that the
corrections are large, one might be lucky, and the
coefficient, effectively, might be much smaller, giving
the behavior found for small $b$ above.
Alternatively, the system might have a local
minimum at some coupling, $g_o$.  (Note that
in any case, the high temperature
calculation is not valid for $T < \Lambda(g)$, at which
point the low energy non-perturbative potential takes
over.)  In this case, one might expect that the
system will roughly track this local minimum,
if it starts out to the left (at stronger coupling),
essentially following adiabatically.  In this case,
one might hope that the system will be gently set
in the true minimum.  We have written toy potentials
with these features, and verified that indeed this
does happen.

These results suggest that
the extent to which the Brustein-Steinhardt problem is a
problem
is quite sensitive to dynamics which are not known as
present, as well as aspects of the initial conditions
of the universe, which are not well understood.

\section{Conclusions}

There has been much discussion through the years of the moduli
problems of string cosmology.  These problems have often
been viewed as providing a serious challenge to the viability of
string theory itself.  The main lesson of the studies here is that
understanding these problems requires both a much better
understanding of moduli stabilization and of the initial
conditions of early universe cosmology than we have at present.
With quite plausible assumptions about how moduli are stabilized,
the moduli need not dominate the energy density of the universe.
With equally plausible assumptions about the conditions of the
early universe, the Brustein-Steinhardt problem is readily
avoided.

We have not offered a complete history of the early universe.
We have not committed ourselves, for example, to the question
of whether we are describing a period before or after inflation,
in part because we suspect that moduli themselves may have
something to do with inflation.  There are many issues which
one would have to consider in a complete model.  For example, the authors
of \cite{riottoetal} have argued that production of non-zero
momentum modes of moduli during inflation, and that this might
even endanger the enhanced symmetry picture we have advocated
here.  On the other hand, as they note, provided the effective
masses of the moduli are of order $H$ during inflation, the
production is modest.  Similarly, given the large mass of the
dilaton, it is likely to decay very early, avoiding the usual
moduli problem.  It is difficult to address these
questions without a detailed model of inflation.
Unfortunately, we have not yet seen a
simple way to fit inflation into this picture, without fine
tuning.  From the results presented here, however, we hope
to have made clear that the moduli problems, as usually
formulated, have plausible, robust solutions.

We know that non-perturbatively
there are many string ground states with unbroken supersymmetry.
It seems likely that, if there are stable string ground states
with broken supersymmetry and vanishing cosmological constant,
there are many of them.  The results reported here
suggest that, as issues involving moduli
stabilization are better understood, cosmology will likely be
crucial to understanding to understanding how the universe finds
itself in the state we see -- and not in one of the myriad other
possibilities.

\noindent
{\bf Acknowledgements:}

\noindent
I thank Y. Nir, Y. Shadmi and Y. Shirman
for their collaboration on earlier parts of this
work, and for many helpful comments.  I also
thank T. Banks
for conversations.   Ref. \cite{steinhardt}
discusses many of the issues raised here,
and I thank P. Steinhardt for
conversations about this work.  This work supported in part by the U.S.
Department of Energy, and by a grant from the U.S.-Israel
Binational Science Foundation.



\begin{thebibliography}{99}
\singlespaced



\bibitem{gauginocondensation}
S. Ferrara, L. Girardello and H.P. Nilles, Phys. Lett. {\bf B155}
(1985) 65; M. Dine, R. Rohm, N. Seiberg and E. Witten, Phys. Lett.
{\bf B156} (1985) 55.

\bibitem{kahlerstabilization}
T. Banks and M. Dine, ``Coping with Strongly Coupled
String Theory,'' Phys.Rev. {\bf D50} (1994), 7454,
hep-th/9406132.

\bibitem{racetrack}
N.V. Krasnikov, Phys. Lett. {\bf B193} (1987) 37; L.J. Dixon,
``Supersymmetry Breaking in String Theory," in {\it The Rice
Meeting:  Proceedings}, B. Bonner and H. Miettinen, eds., World
Scientific (Singapore) 1990;
T.R. Taylor, ``Dilaton, Gaugino Condensation and Supersymmetry
Breaking," Phys. Lett. {\bf 252B} (1990) 59;
B. de Carlos, J.A. Casas and C. Munoz, ``Supersymmetry Breaking and
Determination of the Unification Gauge Coupling Constant
in String Theories,'' Nucl. Phys. {\bf B399} (1993) 623,
hep-th/9204012; V. Kaplunovsky
and J. Louis, ``Phenomenological Aspects of F Theory,'' Phys.
Lett. {\bf B417} (1998) 45, hep-th/9708049.

\bibitem{iy}
K. Izawa and T. Yanagida, ``R Invariant Dilaton Fixing,"
hep-ph/9809366.

\bibitem{dsracetrack}
M. Dine and Y. Shirman, ``Remarks on the Racetrack Scheme,''
hep-th/9906246.


\bibitem{gaillard}
P. Binetruy and M.K. Gaillard, ``Candidates for the Inflaton Field
in Superstring Models,"  Phys.Rev. {\bf D34} (1986) 3069.


\bibitem{banksinflation}
T. Banks, ``Remarks on M Theoretic Cosmology," hep-th/9906126.






\bibitem{ad}
M. Dine and I. Affleck, ``A New Mechanism for Baryogenesis,"  Nucl.
Phys. {\bf B249}, 361 (1985);M. Dine, L. Randall
and S. Thomas, ``Baryogenesis from Flat Directions of the Supersymmetric
Standard Model,"
Nucl.Phys. {\bf B458} (1996), 291, hep-ph/9507453.

\bibitem{bkn}
T. Banks, D. Kaplan and A. Nelson, ``Cosmological Implications of
Dynamical Supersymmetry Breaking,'' Phys. Rev. {\bf D49} (1994)
779, hep-ph/9308292; B. de Carlos, J.A. Casas, F. Quevedo and E.
Roulet, ``Model Independent Properties and Cosmological
Implications of the Dilaton and Moduli Sectors of 4-D Strings,"
Phys.Lett. {\bf B318} (1993) 447,
hep-ph/9308325.


\bibitem{bs}
R. Brustein and P.J. Steinhardt, ``Challenges for Superstring
Cosmology," Phys. Lett. {\bf B302} (19939) 196, hep-th/9212049.


\bibitem{yanagidaetal}
T. Moroi, M. Yamaguchi and T. Yanagida, On the Solution
to the Polonyi Problem with O(10 TeV) Gravitino
Mass in Supergravity,''
 Phys.Lett. {\bf B342} (1995) 105,
hep-ph/9409367.

\bibitem{lateinflation}
L. Randall and S. Thomas, ``Solving the Cosmological
Moduli Problem with Weak Scale Inflation," Nucl. Phys.
{\bf B449} (1995) 229.
hep-ph/9407248.

\bibitem{thermalinflation}
D.H. Lyth and E.D. Stewart, ``Thermal Inflation
and the Moduli Problem,"
Phys.Rev.D53 (1996) 1784,
hep-ph/9510204.



\bibitem{drt}
M. Dine, L. Randall and S. Thomas,
``Supersymmetry Breaking in the Early Universe,"
Phys. Rev. Lett. {\bf 75} (1995), hep-ph/9503303.


\bibitem{dns}
Michael Dine, Yosef Nir and Yael Shadmi
``Enhanced symmetries and the Ground State of String
Theory,'' Phys.Lett. {\bf B438} (1998) 61,
hep-th/9806124.

\bibitem{yukawa}
Michael Dine, ``Seeking the Ground State of String Theory,''
Prog.Theor.Phys.Suppl. {\bf 134} (1999) 1,
hep-th/9903212.



\bibitem{hornemoore}
J. Horne and G. Moore, ``Chaotic Coupling Constants," Nucl. Phys.
{\bf B432} (1994) 105, hep-th/9403058.

\bibitem{banksetal}
T. Banks, M. Berkooz, S.H. Shenker, G. Moore and P.J.
Steinhardt, ``Modular Cosmology," Phys. Rev. {\bf D52} (1995), hep-th/9503114.

\bibitem{largedimensions}
N. Arkani-Hamed, S. Dimopoulos and G. Dvali, ``The Hierarchy
Problem and New Dimensions at a Millimeter,'' Phys. Rev. Lett.
{\bf B429} (1998) 263, hep-ph/9803315; .~Antoniadis,
N.~Arkani-Hamed, S.~Dimopoulos and G.~Dvali, ``New dimensions at a
millimeter to a Fermi and superstrings at a TeV," Phys. Lett. {\bf
B436}, 257 (1998) hep-ph/9804398.  Some precursors of these ideas
can be found in \cite{precursors}.

\bibitem{precursors}
Some precursors of these ideas can be found in V.A.~Rubakov and
M.E.~Shaposhnikov, ``Do We Live Inside A Domain Wall?," Phys.
Lett. {\bf 125B}, 136 (1983); G.~Dvali and M.~Shifman, ``Dynamical
compactification as a mechanism of spontaneous supersymmetry
                  breaking,"
Nucl. Phys. {\bf B504}, 127 (1996) hep-th/9611213; G.~Dvali and
M.~Shifman, ``Domain walls in strongly coupled theories," Phys.
Lett. {\bf B396}, 64 (1997) hep-th/9612128.


\bibitem{rs}
L. Randall and R. Sundrum, ``A Large Mass Hierarchy From a Small
Extra Dimension," hep-ph/9905221; L. Randall and R.
Sundrum, ``An Alternative to
Compactification," hep-th/9906064.



\bibitem{hv}
P. Horava and E. Witten, ``Heterotic and Type I String Dynamics From
Eleven Dimensions'', {\it Nucl. Phys.} {\bf B460}, (1996) 506,
hep-th/9510209.

\bibitem{wittency}
E. Witten, ``Strong Coupling Expansion of Calabi-Yau Compactification,''
hep-th/9602070, Nucl. Phys. {\bf B471} (1996) 135.


\bibitem{steinhardt}
G. Huey, P.J. Steinhardt, Burt A. Ovrut and D. Waldram,
``A Cosmological Mechanism for Stabilizing Moduli,"
hep-th/000112.

\bibitem{riottoetal}
G.F. Guidice, A. Riotto and I. Tkachev, ``Thermal and
Non-Thermal Production of Gravitinos in the Early
Universe," JHEP {\bf 9911} (1999) 036,
hep-ph/9911302.















\bibitem{kapusta}
J. I. Kapusta,  {\it Finite Temperature Field Theory},
Cambridge University Press, Cambridge 1989.






\end{thebibliography}
\end{document}